\documentclass[prd,twocolumn] {revtex4}
\usepackage[latin1]{inputenc}
\usepackage{graphicx}
\usepackage{epsfig}
\usepackage{bm}
\usepackage{amssymb}
\usepackage{float}
\usepackage{amsmath}
\usepackage{dcolumn}
\usepackage{cancel}
\usepackage{color}
\usepackage{epstopdf}
\usepackage{nccbbb}
\usepackage{lipsum}

\begin{document}

\title {A new unified dark sector model and its implications on the $\sigma_8$ and $S_8$ tensions}

\author{Yan-Hong Yao}
\email{yaoyh29@mail.sysu.edu.cn}
\author{Jian-Qi Liu}
\author{Zhi-Qi Huang}
\author{Jun-Chao Wang}
\author{Yan Su}

\affiliation{School of Physics and Astronomy, Sun Yat-sen University,
2 Daxue Road, Tangjia, Zhuhai, People's Republic of China}

\begin{abstract}
In this paper, we introduced the Unified Three-Form Dark Sector (UTFDS) model, a unified dark sector model that combines dark energy and dark matter through a three-form field. In this framework, the potential of the three-form field acts as dark matter, while the kinetic term represents dark energy. The interaction between dark matter and dark energy is driven by the energy exchange between these two terms. Given the dynamical equations of UTFDS, we provide an autonomous system of evolution equations for UTFDS and perform a stability analysis of its fixed points. The result aligns with our expectations for a unified dark sector. Furthermore, we discover that the dual Lagrangian of the UTFDS Lagrangian is equivalent to a Dirac-Born-Infeld (DBI) Lagrangian. By fixing the parameter $\kappa X_0$ to 250, 500, 750, we refer to the resulting models as the $\overline{\rm UTFDS}$ model with $\kappa X_0$=250, 500, 750, respectively. We then place constraints on these three $\overline{\rm UTFDS}$ models and the $\Lambda$CDM model in light of the Planck 2018 Cosmic Microwave Background (CMB) anisotropies, Redshift Space Distortions (RSD) observations, Baryon Acoustic Oscillation (BAO) measurements, and the $S_8$ prior chosen according to the KiDS1000 Weak gravitational Lensing (WL) measuement. We find that the $\overline{\rm UTFDS}$ model with $\kappa X_0$=500 is the only one among the four models where both $\sigma_8$ and $S_8$ tensions, between CMB and RSD+BAO+WL datasets, are below 2.0$\sigma$. Furthermore, the tensions are relieved without exacerbating the $H_0$ tension. Although both the CMB and RSD+BAO+WL datasets provide definite/positive evidence favoring $\Lambda$CDM over the $\overline{\rm UTFDS}$ model with $\kappa X_0$=500, the evidence is not strong enough to rule out further study of this model.

\textbf{}
\end{abstract}

\maketitle

\section{Introduction}
\label{sec:0}
The Lambda Cold Dark Matter ($\Lambda$CDM) model has been widely accepted as the standard cosmological model for over 20 years and has been validated by a wide range of astronomical observations across different scales~\cite{riess1998observational,perlmutter1999measurements,dunkley2011atacama,hinshaw2013nine,story2015measurement,alam2017clustering,troxel2018dark,aghanim2020planck1}. However, with the continuous advancement of observational techniques and the more precise determination of cosmological parameters, discrepancies between cosmological parameters derived from different types of observations have started to emerge, which were previously unnoticed.

One of the most prominent discrepancies is the $H_0$ tension~\cite{riess2022comprehensive,aghanim2020planck1}, which refers to about $5\sigma$ tension between late- and early-time independent measurements of the Hubble constant. Whether such tension requires new physics, and what this new physics might be, has become the subject of an ongoing and rapidly evolving research direction~\cite{di2021realm,reeves2023restoring,vagnozzi2021consistency,poulin2019early,agrawal2019rock,smith2020oscillating,lin2019acoustic,niedermann2021new,freese2021chain,ye2020hubble,akarsu2020graduated,braglia2020unified,vagnozzi2018constraints,visinelli2019revisiting,huang2016dark,vagnozzi2020new,martinelli2019cmb,alestas2020h,d2021limits,yang2019observational,alestas2021late,battye2014evidence,Zhang2014Neutrinos,zhang2015sterile,feng2018searching,zhao2018measuring,choudhury2019constraining,yao2023can,yao2021relieve,yao2020new1,kumar2016probing,kumar2017echo,kumar2019dark,nunes2022new,yang2018tale,di2017can,yang2018interacting,di2020nonminimal,cheng2020testing,lucca2020tensions,gomez2020update,yang2019dark,yang2020dynamical}. Aside from the Hubble tension, the $\sigma_8$ tension, or $S_8=\sigma_8\sqrt{\Omega_{\rm m}/0.3}$ tension~\cite{macaulay2013lower,joudaki2016cfhtlens,bull2016beyond,joudaki2017kids,nesseris2017tension,kazantzidis2018evolution,asgari2020kids+,hildebrandt2020kids+,skara2020tension,abbott2020dark,joudaki2020kids+,heymans2021kids,asgari2021kids,loureiro2021kids,abbott2022dark,amon2022dark,secco2022dark,philcox2022boss}, where $\Omega_{\rm m}$ is the matter density parameter and $\sigma_8$ is the matter fluctuation amplitude on scales of $ 8h^{-1}{\rm Mpc}$, is a milder yet not less enduring discrepancy. The $\sigma_8$ (or $S_8$) tension indicates that the Universe predicted by low-redshift observations, such as Weak gravitational Lensing (WL) and Galaxy Clustering (GC), is smoother than the one predicted by Cosmic Microwave Background (CMB) observations. Quantitatively, such tension has reached a level of $2$-$3\sigma$. Although the status of the $\sigma_8$ (or $S_8$) tension may be less certain compared to the Hubble tension, it is undeniable that there is some disagreement between low- and high-redshift measurements of matter fluctuations. It is therefore worth exploring whether new physics could resolve or at least reduce this tension. To achieve this, many models have been proposed, including ultra-light axions~\cite{rogers2023ultra}, decaying dark matter~\cite{enqvist2015decaying,di2018reducing,pandey2020alleviating}, dark energy and dark matter interaction model~\cite{di2020interacting}, modified gravity models~\cite{dossett2015constraints,kazantzidis2021sigma,marra2021rapid}, and unified dark sector models~\cite{kamenshchik2001alternative,bertacca2010unified,camera2019does,wang2024pagelikeunifieddarkfluid}. It is important to note that while most of these models can alleviate the $\sigma_8$ (or $S_8$) tension to some degree, they often do so at the cost of exacerbating the $H_0$ tension.

In this article, we propose a novel unified dark sector model, termed the Unified Three-Form Dark Sector (UTFDS) model, in which dark energy and dark matter are unified by using a three-form field, where the potential of the three-form field represents dark matter and the kinetic represents dark energy. The energy transformation between potential and kinetic corresponds to the interaction between dark matter and dark energy. Decomposing UTFDS into dark energy and dark matter from this perspective, we find that the resulting dark matter can have a non-zero sound speed, therefore, by fixing the value of a specific parameter in this model, we can regulate the magnitude of the dark matter sound speed, offering an alternative approach to addressing the $\sigma_8$ (or $S_8$) tension. We refer to the class of new models obtained by fixing the value of this
particular parameter in the UTFDS model as the $\overline{\rm UTFDS}$ models.

The rest of this paper is organized as follows. In section~\ref{sec:1}, we present the dynamical equations of UTFDS at both the background and linear levels. Additionally, we derive the autonomous system of evolution equations for UTFDS and conduct a stability analysis of its fixed points. Furthermore, we derive the dual Lagrangian of the UTFDS Lagrangian in the end of this section. In section~\ref{sec:2}, we describe the observational datasets and the statistical methodology. In section~\ref{sec:3}, we report the results within three particular $\overline{\rm UTFDS}$ models. The last section concludes.
\section{theoretical framework of UTFDS}
\label{sec:1}
In this section, we present the UTFDS model, in which dark energy and dark matter are unified by using a canonical three-form field minimally coupled to Einstein gravity. The Lagrangian is
\begin{equation}\label{equ:lglr}
\mathcal{L}=\frac{R}{2\kappa^2}+K-V(y)+\mathcal{L}_{b}+\mathcal{L}_{r},
\end{equation}
where $R$ denotes the Ricci scalar and $\kappa=\sqrt{8\pi G}$ is the inverse of the reduced Planck mass. Subscripts $b$ and $r$ denote baryon and radiation, respectively. And $K = -\frac{1}{48}F^2$, $y = \frac{1}{12}A^2$, here $A$ and $F=dA$ represent the three-form field and the field strength tensor, respectively. For the Lagrangian given in Eq.~\ref{equ:lglr}, applying the principle of least action gives the Einstein field equation:
\begin{equation}\label{}
  R_{\mu\nu}-\frac{1}{2}g_{\mu\nu}R=\kappa^2 (T_{\mu\nu}^{(A)}+T_{\mu\nu}^{(b)}+T_{\mu\nu}^{(r)}),
\end{equation}
where $R_{\mu\nu}$ is the Ricci tensor, and
\begin{equation}\label{}
\begin{split}
 T_{\mu\nu}^{(A)}=&\frac{1}{6}F_{\mu\alpha\beta\gamma}F_{\nu}^{\alpha\beta\gamma}+\frac{1}{2}\frac{dV}{dy}A_{\mu}^{\alpha\beta}A_{\nu\alpha\beta}+g_{\mu\nu}(K-V),
\end{split}
\end{equation}
By varying the action corresponding to the above Lagrangian with respect to the three-form field, we have the following equation of motion,
\begin{equation}\label{eqa:dlxfc}
 \nabla_{\alpha}(F^{\alpha\mu\nu\rho})=\frac{dV}{dy}A^{\mu\nu\rho}.
\end{equation}
A canonical three-form field is equivalent to an effective perfect fluid characterized by the following energy-momentum tensor~\cite{PhysRevD.96.023516}:
\begin{equation}\label{}
\begin{split}
 T_{\mu\nu}^{A}=&(\rho_{A}+p_{A})u_{\mu}^{(A)}u_{\nu}^{(A)}+p_{A}g_{\mu\nu},
\end{split}
\end{equation}
where
\begin{eqnarray}
  \rho_{A} &=& K+V,\\
  p_{A} &=&-K-V+2y\frac{dV}{dy},  \\
  u^{(A)\mu} &=& \frac{\epsilon^{\mu\alpha\beta\gamma}A_{\alpha\beta\gamma}}{3!\sqrt{2y}}.
\end{eqnarray}
The equation of motion of three-form is equivalent to the following equation:
\begin{equation}\label{}
  \nabla_{\mu}((\rho_{A}+p_{A})u_{\mu}^{(A)}u_{\nu}^{(A)}+p_{A}g_{\mu\nu})=0.
\end{equation}
It is important to note that, to extract the values of the matter density parameter and the matter fluctuation amplitude from various astronomical data, we need to decompose UTFDS into dark energy and dark matter. There are infinitely many ways to do this; for simplicity, we adopt the following decomposition method:
\begin{eqnarray}
  \rho_{\rm de} &=& K,\\
  \rho_{\rm dm} &=& V,\\
  p_{\rm de} &=&-K,\\
  p_{\rm dm} &=&-V+2y\frac{dV}{dy},\\
  u^{(\rm dm)\mu} &=&  u^{(A)\mu},
\end{eqnarray}
that is to say, we assign the portion of the physical quantities associated with kinetic energy to dark energy, while the remaining part is assigned to dark matter, as we mentioned in the introduction. By adopting this decomposition method, the energy-momentum tensor can be rewritten as:
\begin{equation}\label{}
\begin{split}
 T_{\mu\nu}^{(A)}=&p_{\rm de}g_{\mu\nu}+(\rho_{\rm dm}+p_{\rm dm})u_{\mu}^{(\rm dm)}u_{\nu}^{(\rm dm)}+p_{\rm dm}g_{\mu\nu},
\end{split}
\end{equation}
and the continuity equation for dark matter can be written as:
\begin{equation}\label{}
  \nabla_{\mu}((\rho_{\rm dm}+p_{\rm dm})u_{\mu}^{(\rm dm)}u_{\nu}^{(\rm dm)}+p_{\rm dm}g_{\mu\nu})=-\nabla_{\mu}(p_{\rm de}g_{\mu\nu}),
\end{equation}
we can see that dark matter and dark energy are not conserved individually, which suggests that there is an interaction between them.
\subsection{Background cosmology}
The homogeneous, isotropic, and spatially flat space-time is described by the Friedmann-Robertson-Walker (FRW) metric,
\begin{equation}\label{}
  ds^{2}=-dt^{2}+a(t)^{2}d\vec{x}^{2},
\end{equation}
here $a(t)$ is the scale factor, and the three-form field is assumed as a time-like component of a dual vector field in order to be compatible with FRW symmetries~\cite{Koivisto2009Inflation2}
\begin{equation}
  A_{ijk} =a^3 \varepsilon_{ijk} X,
\end{equation}
then we have the following Friedmann equations:
\begin{eqnarray}
  H^{2} &=& \frac{\kappa^{2}}{3}(\bar{\rho}_{A}+\bar{\rho}_{b}+\bar{\rho}_{r}),\\
 \dot{H} &=&-\frac{\kappa^{2}}{2}(\bar{\rho}_{A}+\bar{p}_{A}+\bar{\rho}_{b}+\frac{4}{3}\bar{\rho}_{r}),
\end{eqnarray}
where
\begin{eqnarray}
\bar{\rho}_{A} &=&\frac{1}{2}(\dot{X}+3HX)^{2}+V,\\
\bar{p}_{A} &=&-\frac{1}{2}(\dot{X}+3HX)^{2}-V+V_{,X} X,
\end{eqnarray}
and the independent equation of motion of the three-form field is
\begin{equation}\label{}
\begin{split}
   &\ddot{X}+3\dot{H}X+3H\dot{X}+V_{,X}=0,
\end{split}
\end{equation}
the above equation of motion of three-form is equivalent to the following continuity equation for the dark sector:
\begin{equation}
 \dot{\bar{\rho}}_{A}+3H\bar{\rho}_{A}(1+w_{A}) =0,
\end{equation}
where the Equation of State (EoS) $w_{A}$ is written as:
\begin{equation}\label{}
  w_{A}=\frac{\bar{p}_{A}}{\bar{\rho}_{A}}=-1 +\frac{V_{,X}X}{\bar{\rho}_{A}},
\end{equation}
by applying the decomposition method mentioned earlier to the dark sector, we obtain the energy density, pressure, and EoS for dark matter and dark energy as follows:
\begin{eqnarray}
\bar{\rho}_{\rm de} &=&\frac{1}{2}(\dot{X}+3HX)^{2},\\
\bar{p}_{\rm de} &=&-\frac{1}{2}(\dot{X}+3HX)^{2},\\
\bar{\rho}_{\rm dm} &=&V,\\
\bar{p}_{\rm dm} &=&-V+V_{,X} X,
\end{eqnarray}
\begin{equation}\label{}
  w_{\rm de}=\frac{\bar{p}_{\rm de}}{\bar{\rho}_{\rm de}}=-1,
\end{equation}
\begin{equation}\label{equ:awzztfc}
  w_{\rm dm}=\frac{\bar{p}_{\rm dm}}{\bar{\rho}_{\rm dm}}=-1 +\frac{V_{,X}X}{V}.
\end{equation}
We can see that dark energy has the same EoS as vacuum energy. However, it is important to note that its energy density varies over time due to the interaction between dark energy and dark matter. To obtain the specific form of the EoS for dark matter, we must first determine the form of the potential of the three-form field. For the purpose of phenomenological study, we set the potential of the three-form as follows:
\begin{equation}\label{}
  V =\widetilde{V}(1+\frac{\kappa^{2}}{6}A^{2})^{\frac{1}{2}}=\widetilde{V} \sqrt{1+\kappa^2X^2},
\end{equation}
by substituting the EoS for dark matter into the Eq.~\ref{equ:awzztfc}, we find that when $X\gg1$ , dark matter behaves like cold dark matter at the background level.

\begin{figure*}
	\centering
	\includegraphics[scale=0.6]{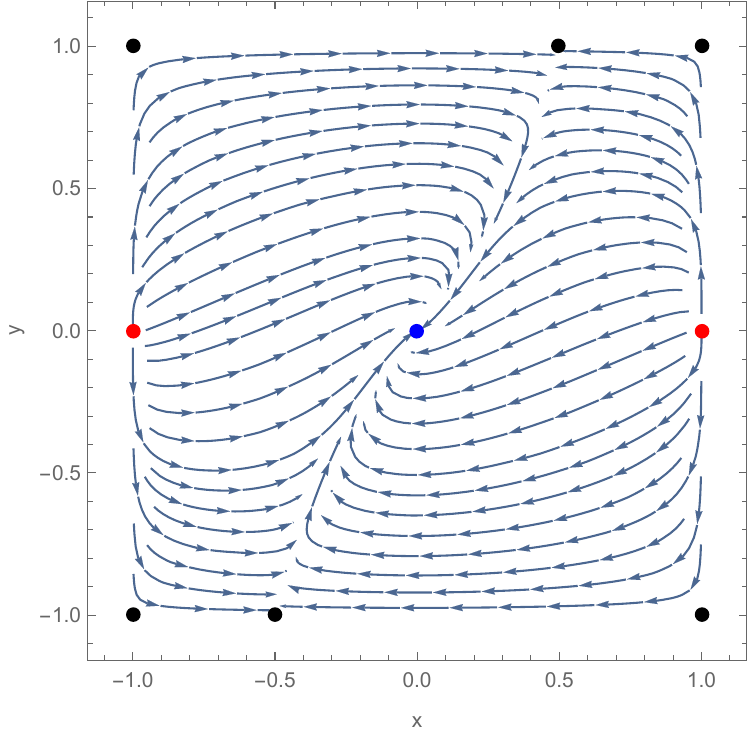}
	\caption{Phase space trajectory diagram for the autonomous system under the assumption that $w$=$v$=0. Here the blue point represents the attractor, two red points represent two repellers, six black points represent six saddle points. }\label{fig:xkj}
\end{figure*}

In order to study the dynamical behaviors of the UTFDS model, it is convenient to introduce the following dimensionless variables:
\begin{widetext}
\begin{equation}\label{}
 x=\frac{2}{\pi}\arctan{\frac{3\kappa X}{\sqrt{6}}},\hspace{1cm}y=\frac{\kappa}{\sqrt{6}}(X^{\prime}+3X),\hspace{1cm}z=\frac{\kappa\sqrt{V}}{\sqrt{3}H},\hspace{1cm}w=\frac{\kappa\sqrt{\bar{\rho}_{b}}}{\sqrt{3}H},\hspace{1cm}v=\frac{\kappa\sqrt{\bar{\rho}_{r}}}{\sqrt{3}H},
\end{equation}
\end{widetext}
The autonomous system of evolution equations then can be written as follows by applying the Friedmann equations and equation of motion,
\begin{widetext}
\begin{eqnarray}
  x^{\prime} &=& \frac{6}{\pi}\cos^2{\frac{\pi x}{2}}(y-\tan{\frac{\pi x}{2}}),\\
  y^{\prime} &=&\frac{\tan{\frac{\pi x}{2}}}{1+\frac{2}{3}\tan^2{\frac{\pi x}{2}}}(1-y^2-w^2-v^2)(y\tan{\frac{\pi x}{2}}-1)+\frac{3}{2}(w^2+\frac{4}{3}v^2)y,\\
  w^{\prime} &=&-\frac{3}{2}w+\frac{3}{2}w(w^2+\frac{4}{3}v^2+\frac{2}{3}\frac{\tan^2{\frac{\pi x}{2}}}{1+\frac{2}{3}\tan^2{\frac{\pi x}{2}}}(1-y^2-w^2-v^2)),\\
  v^{\prime} &=&-2v+\frac{3}{2}v(w^2+\frac{4}{3}v^2+\frac{2}{3}\frac{\tan^2{\frac{\pi x}{2}}}{1+\frac{2}{3}\tan^2{\frac{\pi x}{2}}}(1-y^2-w^2-v^2)),
\end{eqnarray}
\end{widetext}
the prime here stands for the derivative with respect to e-folding time $N=\ln a$. We note that $z$ has been eliminated by the Friedmann constraint written in terms of these variables
$y^2+z^2+w^2+v^2=1$. Since the above autonomous system has no analytical solution and involves four variables, making it impossible to plot a phase space trajectory diagram, we assume here that baryons and radiation can be neglected, i.e. $w$=$v$=0. Therefore, the new autonomous system can be written as:
\begin{eqnarray}
  x^{\prime} &=& \frac{6}{\pi}\cos^2{\frac{\pi x}{2}}(y-\tan{\frac{\pi x}{2}}),\\
  y^{\prime} &=&\frac{\tan{\frac{\pi x}{2}}}{1+\frac{2}{3}\tan^2{\frac{\pi x}{2}}}(1-y^2)(y\tan{\frac{\pi x}{2}}-1).
\end{eqnarray}
There are nine fixed points for this autonomous system of evolution equations, which are presented in Tab.~\ref{tab:gdd}. The values of $\Omega_{\rm de}$, $\Omega_{\rm dm}$, and $w_{A}$ in this table can be calculated by using the dimensionless variables through the following equations:
\begin{eqnarray}\label{}
  \Omega_{\rm de}&=&y^2,\\
  \Omega_{\rm dm}&=&1-y^2-w^2-v^2,\\
   w_{A}&=&-1+\frac{1-y^2-w^2-v^2}{1-w^2-v^2}\frac{\frac{2}{3}\tan^2{\frac{\pi x}{2}}}{1+\frac{2}{3}\tan^2{\frac{\pi x}{2}}},
\end{eqnarray}
The trajectories ($x(N)$,$y(N)$) with a wide range of initial conditions can be visualized by Fig.~\ref{fig:xkj}. As is shown in the Tab.~\ref{tab:gdd} and Fig.~\ref{fig:xkj}, the trajectories begin in a matter-dominated universe, move along the de Sitter saddle point, and eventually approach the de Sitter attractor. Therefore, UTFDS indeed behaves like dust matter at high redshifts and like a cosmological constant at low redshifts, at least at the background level, which meets our requirements for a unified dark sector.
\begin{table*}[ht]
    \centering
    \scalebox{1}[1]{
    \begin{tabular}{|c|c|c|c|c|c|c|c|c|c|}
        \hline
                &  $a$ &  $b$ &  $c$& $d$ &  $e$ &  $f$ &  $g$ & $ h$  &  $ i $ \\ \hline
        $(x,y)$ & ~$(0,0)$~ & ~$(1,0)$~ &~$(-1,0)$~ & ~(1,1)~ & ~(1,$-1$)~ & ~($-1$,1)~ & ~($-1$,$-1$)~ & ~($-\frac{1}{2}$,$-1$)~ &~($\frac{1}{2}$,1)~\\ \hline \hline
         type  &attractor  & repeller  & repeller    & saddle point  & saddle point  &  saddle point   &   saddle point&   saddle point&   saddle point    \\
        $ \Omega_{\rm de} $ &0 & 0   &  0    &  1   & 1  &  1  &   1  &   1  &  1    \\
        $ \Omega_{\rm dm} $ &1 &  1  &  1    &  0   & 0  &  0  &  0   &   0   &  0   \\
        $  w_{A}      $ &$-1$ & 0 & 0 & $-1$ & $-1$ & $-1$ & $-1$ &  $-1$ & $-1$  \\
        \hline
    \end{tabular}}
    \caption{Fixed points of the autonomous system under the assumption that $w$=$v$=0.}
    \label{tab:gdd}
\end{table*}

Finally, we write down the dimensionless Hubble parameter that will be used when constraining the model with observational data,
\begin{equation}\label{}
  E^2=\frac{H^2}{H_0^2}=\frac{z_0^2\sqrt{\frac{1+\frac{2}{3}\tan^2{\frac{\pi x}{2}}}{1+\frac{2}{3}\tan^2{\frac{\pi x_0}{2}}}}+w_0^2(\frac{a_0}{a})^3+v_0^2(\frac{a_0}{a})^4}{1-y^2},
\end{equation}
here and in the following, the subscript 0 denotes the present moment.
\subsection{Cosmological perturbations}
In the conformal Newtonian gauge, the perturbed FRW metric takes the form
\begin{equation}\label{}
  ds^{2}=-(1+2\psi)dt^{2}+a(t)^2(1-2\phi)d\vec{x}^{2},
\end{equation}
where $\psi$ and $\phi$ are the metric potentials and $\vec{x}$ represents the three spatial coordinates, and we parametrize the scalar fluctuations of the three-form using the two scalars $\alpha$ and $\alpha_0$~\cite{Koivisto2009Inflation2}:
\begin{eqnarray}
  A_{0ij} &=&a^2 \varepsilon_{ijk} \alpha_{,k},\\
  A_{ijk} &=&a^3 \varepsilon_{ijk} (X+\alpha_0).
\end{eqnarray}
The perturbed Einstein's equation reads:
\begin{equation}\label{}
  \delta R_{\mu\nu}-\frac{1}{2}\delta g_{\mu\nu}R-\frac{1}{2}g_{\mu\nu}\delta R=\kappa^2 (\delta T_{\mu\nu}^{(A)}+\delta T_{\mu\nu}^{(b)}+\delta T_{\mu\nu}^{(r)}),
\end{equation}
and the perturbed components of the energy-momentum tensor read:
\begin{widetext}
\begin{eqnarray}
-\delta T_{0}^{(A)0}&=&\delta\rho_{A}=(\dot{X}+3HX)(\dot{\alpha}_0+3H\alpha_0+(\dot{X}+3HX)(3\phi-\psi)-\frac{ \nabla^2}{a^2}\alpha)+V_{,X}(\alpha_0+3X\phi),\\
\delta T_{i}^{(A)0}&=&V_{,X}X\delta u^{(A)}_{i}=-V_{,X}\alpha_{,i},\\
\delta T_{i}^{(A)i}&=&\delta p_{A}=-(\dot{X}+3HX)(\dot{\alpha}_0+3H\alpha_0+(\dot{X}+3HX)(3\phi-\psi)-\frac{ \nabla^2}{a^2}\alpha)+V_{,XX}X(\alpha_0+3X\phi),\\
\delta T_{j}^{(A)i}&=&0.
\end{eqnarray}
\end{widetext}
The equation of motion of the three-form field yields the equation of motion for $\alpha_0$:
\begin{widetext}
\begin{eqnarray}\label{}
&&\ddot{\alpha}_0+3\dot{(H\alpha_0)}+(\ddot{X}+3\dot{(HX)})(3\phi-2\psi)+(\dot{X}+3HX)(3\dot{\phi}-\dot{\psi})+\frac{\nabla^2}{a^2}(2H\alpha-\dot{\alpha})
  +V_{,XX}(\alpha_0+3X\phi)=0,\\
 &&\dot{\alpha}_0+3H\alpha_0+(\dot{X}+3HX)(3\phi-\psi)-\frac{ \nabla^2}{a^2}\alpha+\frac{V_{,X}\alpha}{X}=0,
\end{eqnarray}
\end{widetext}
the sound speed squared of the dark sector in its rest frame is
\begin{equation}\label{equ:azfss}
  c_{(A)s}^2=\frac{\delta p_{A}}{\delta \rho_{A}}\mid_{\alpha=0}=\frac{V_{,XX}X}{V_{,X}}=\frac{1}{1+\frac{2}{3}\tan^2{\frac{\pi x}{2}}}.
\end{equation}
By defining the relative density and velocity divergence perturbations of the dark sector as follows:
\begin{equation}\label{}
  \delta_{A}=\frac{\delta \rho_{A}}{\bar{\rho}_{A}},
\end{equation}
\begin{equation}
  \theta_{A}=a\partial_i\delta u^{(A)i},
\end{equation}
we can then derive the following continuity and Euler equations (in the Fourier space) for effective perfect fluids from the condition that the divergence of the UTFDS energy-momentum tensor vanishes, i.e. $\nabla^\mu T_{\mu\nu}^{(A)}=0$, this system of equations is equivalent to the equation of motion for $\alpha_0$,
\begin{widetext}
\begin{eqnarray}\label{}
&&\dot{\delta}_{A}+3H(c_{(A)s}^2-w_{A})\delta_{A}+(1+w_{A})\frac{\theta_{A}}{a}+3H(3H(1+w_{A})(c_{(A)s}^2-w_{A})+\dot{w}_{A})\frac{a\theta_{A}}{k^2}-3(1+w_{A})\dot{\phi}=0,\\
&& \frac{\dot{\theta}_{A}}{a}+H(1-3c_{(A)s}^2)\frac{\theta_{A}}{a}-\frac{c_{(A)s}^2}{1+w_{A}}\frac{k^2}{a^2}\delta_{A}-\frac{k^2}{a^2}\psi=0.
\end{eqnarray}
\end{widetext}
By substituting the Eq.~42 and Eq.~\ref{equ:azfss} into the above system of equations, we obtain the following system of equations:
\begin{widetext}
\begin{eqnarray}\label{}
\begin{split}
&\dot{\delta}_{A}+3H(\frac{2(1-w^2-v^2)+\frac{2}{3}y^2\tan^2{\frac{\pi x}{2}}}{(1-w^2-v^2)(1+\frac{2}{3}\tan^2{\frac{\pi x}{2}})})\delta_{A}+\frac{1-y^2-w^2-v^2}{1-w^2-v^2}\frac{\frac{2}{3}\tan^2{\frac{\pi x}{2}}}{1+\frac{2}{3}\tan^2{\frac{\pi x}{2}}}\frac{\theta_{A}}{a}+6H^2\frac{1-y^2-w^2-v^2}{1-w^2-v^2}\times\\&\frac{y\tan{\frac{\pi x}{2}}(2+\frac{2}{3}\tan^2{\frac{\pi x}{2}})}{(1+\frac{2}{3}\tan^2{\frac{\pi x}{2}})^2}\frac{a\theta_{A}}{k^2}-\frac{1-y^2-w^2-v^2}{1-w^2-v^2}\frac{2\tan^2{\frac{\pi x}{2}}}{1+\frac{2}{3}\tan^2{\frac{\pi x}{2}}}\dot{\phi}
  =0,\\
&\frac{\dot{\theta}_{A}}{a}+H(\frac{\frac{2}{3}\tan^2{\frac{\pi x}{2}}-2}{1+\frac{2}{3}\tan^2{\frac{\pi x}{2}}})\frac{\theta_{A}}{a}-\frac{3(1-w^2-v^2)}{2(1-y^2-w^2-v^2)\tan^2{\frac{\pi x}{2}}}\frac{k^2}{a^2}\delta_{A}
 -\frac{k^2}{a^2}\psi=0.
\end{split}
\end{eqnarray}
\end{widetext}
In practice, we can treat the dark sector as a whole, then solve the perturbed Einstein's equation and all other perturbation equations to obtain relative density and velocity divergence perturbations of dark sector. Afterward, we express the relative density and velocity divergence perturbations of dark matter, as well as relative density perturbations of dark energy, in terms of relative density and velocity divergence perturbations of dark sector through linear combinations. Now let's derive these linear representations.

By applying our decomposition method to the dark sector's perturbations, we obtain the energy density perturbations and pressure perturbations for dark matter and dark energy as follows:
\begin{widetext}
\begin{eqnarray}
\delta\rho_{\rm de}&=&(\dot{X}+3HX)(\dot{\alpha}_0+3H\alpha_0+(\dot{X}+3HX)(3\phi-\psi)-\frac{ \nabla^2}{a^2}\alpha),\hspace{1cm}\delta\rho_{\rm dm}=V_{,X}(\alpha_0+3X\phi),\\
\delta p_{\rm de}&=&-(\dot{X}+3HX)(\dot{\alpha}_0+3H\alpha_0+(\dot{X}+3HX)(3\phi-\psi)-\frac{ \nabla^2}{a^2}\alpha),\hspace{1cm}\delta p_{\rm dm}=V_{,XX}X(\alpha_0+3X\phi),
\end{eqnarray}
\end{widetext}
given the above equations, we find that the sound speed squared of dark matter is equal to the sound speed squared of the dark sector, i.e.
\begin{equation}\label{}
  c_{(\rm dm)s}^2=\frac{\delta p_{\rm dm}}{\delta \rho_{\rm dm}}\mid_{\alpha=0}=\frac{1}{1+\frac{2}{3}\tan^2{\frac{\pi x}{2}}}.
\end{equation}
By using Eq.~48 and Eq.~53, we obtain the following equation:
\begin{equation}\label{}
  \alpha_0+3X\phi=\frac{\bar{\rho}_{A}}{V_{,X}}(\delta_{A}+\frac{3(1+w_{A})Hy}{\tan{\frac{\pi x}{2}}}\frac{a\theta_{A}}{k^2}).
\end{equation}
Substituting this equation into Eq.~61, we obtain:
\begin{equation}
\begin{split}
\delta_{\rm dm}&=\frac{\delta\rho_{\rm dm}}{\bar{\rho}_{\rm dm}}=\frac{\bar{\rho}_{A}}{V}(\delta_{A}+\frac{3(1+w_{A})Hy}{\tan{\frac{\pi x}{2}}}\frac{a\theta_{A}}{k^2})\\&=\frac{1-w^2-v^2}{1-y^2-w^2-v^2}\delta_{A}+\frac{2yH\tan{\frac{\pi x}{2}}}{1+\frac{2}{3}\tan^2{\frac{\pi x}{2}}}\frac{a\theta_{A}}{k^2},
\end{split}
\end{equation}
considering that the definition of the relative density perturbation of the dark sector is expressed as follows:
\begin{equation}
\delta_{A}=\frac{\delta\rho_{\rm de}+\delta\rho_{\rm dm}}{\bar{\rho}_{A}}.
\end{equation}
We obtain the following relative density perturbation of dark energy:
\begin{equation}
\delta_{\rm de}=\frac{\delta\rho_{\rm de}}{\bar{\rho}_{\rm de}}=-\frac{2H(1-y^2-w^2-v^2)\tan{\frac{\pi x}{2}}}{y(1+\frac{2}{3}\tan^2{\frac{\pi x}{2}})}\frac{a\theta_{A}}{k^2}.
\end{equation}
Finally, since dark matter and the dark sector have the same four-velocity, they naturally share the same velocity divergence perturbation, which means
\begin{equation}\label{}
  \theta_{\rm dm}=\theta_{A}.
\end{equation}
At this point, we have provided all the linear representations.
\subsection{The dual Lagrangian}
In this subsection, we derive the dual Lagrangian of the UTFDS Lagrangian. To begin, we introduce a parent Lagrangian that is equivalent to the UTFDS Lagrangian, as follows:
\begin{equation}\label{}
  \mathcal{L}_p=\frac{1}{48}F^2-\frac{1}{6}A\nabla\cdot F-V(A^2),
\end{equation}
the Hodge dual of the three-form is a vector $(\ast A)$. Writing the parent Lagrangian in terms of the dual forms
\begin{equation}\label{equ:dual}
  F=\epsilon(\ast F)=\epsilon \phi, A=\epsilon(\ast A)=\epsilon B,
\end{equation}
where $\phi$ is a scalar field and $B$ is a vector. Substitute Eq.~\ref{equ:dual} into the parent Lagrangian, we obtain
\begin{equation}\label{}
  \mathcal{L}=-\frac{1}{2}\phi^2-B\cdot\nabla\phi-V(-6B^2),
\end{equation}
now, we can integrate the vector from parent Lagrangian above to obtain a scalar field theory. In order to do that, we first derive the Euler-Lagrange equation for $B$ as follows:
\begin{equation}\label{equ:elefb}
  \nabla\phi=12V^{\prime}B,
\end{equation}
the prime here stands for the derivative with respect to $A^2$, and the Eq.~\ref{equ:elefb} implies:
\begin{equation}\label{}
  (\nabla\phi)^2=144(V^{\prime})^2B^2,
\end{equation}
provided the specific expressions for $V$ and $V^{\prime}$ in term of $B$ as follows,
\begin{equation}\label{}
  V=V_0\sqrt{1+\frac{\kappa^{2}}{6}A^{2}}=V_0\sqrt{1-\kappa^{2}B^{2}},
\end{equation}
\begin{equation}\label{}
  V^{\prime}=\frac{dV}{dA^2}=\frac{\kappa V_0}{12\sqrt{1+\frac{\kappa^{2}}{6}A^{2}}}=\frac{\kappa^2 V_0}{12\sqrt{1-\kappa^{2}B^{2}}},
\end{equation}
the Eq.~\ref{equ:elefb} can be rewritten as:
\begin{equation}\label{}
  \nabla\phi=\frac{\kappa^2 V_0B}{\sqrt{1-\kappa^{2}B^{2}}},
\end{equation}
which leads to:
\begin{equation}\label{}
  (\nabla\phi)^2=\frac{\kappa^4 V_0^2B^2}{1-\kappa^{2}B^{2}}\rightarrow\sqrt{1+\frac{(\nabla\phi)^2}{\kappa^2V_0^2}}=\frac{1}{\sqrt{1-\kappa^{2}B^{2}}},
\end{equation}
therefore, we obtain the dual Lagrangian, in term of scalar field, of the UTFDS Lagrangian as follows:
\begin{equation}\label{equ:dolsl}
\begin{split}
  \mathcal{L}=&-\frac{1}{2}\phi^2-B\cdot\nabla\phi-V(-6B^2)=-\frac{1}{2}\phi^2+\frac{ V_0}{\sqrt{1-\kappa^{2}B^{2}}}\\&=-\frac{1}{2}\phi^2+V_0\sqrt{1+\frac{(\nabla\phi)^2}{\kappa^2V_0^2}}.
\end{split}
\end{equation}
Interestingly, if we apply transformations $\varphi=\frac{\phi}{\kappa\sqrt{V_0}}$ and $\mathcal{\tilde{L}}=-\mathcal{L}$, we can obtain a new Lagrangian that is equivalent to Eq.~\ref{equ:dolsl} and in term of scalar field $\varphi$,
\begin{equation}\label{equ:dolsl2}
  \mathcal{\tilde{L}}=-V_0\sqrt{1+\frac{(\nabla\varphi)^2}{V_0}}+V_0-V_0(1-\frac{1}{2}\kappa^2\varphi^2).
\end{equation}
Defining the following two functions,
\begin{eqnarray}
   \mathcal{F}&=& \frac{1}{V_0}, \\
  \mathcal{V}(\varphi) &=& V_0(1-\frac{1}{2}\kappa^2\varphi^2),
\end{eqnarray}
the Lagrangian Eq.~\ref{equ:dolsl2} can be rewritten as the following form:
\begin{equation}\label{}
  \mathcal{\tilde{L}}=-\mathcal{F}(\varphi)^{-1}\sqrt{1+\mathcal{F}(\varphi)(\nabla\varphi)^2}+\mathcal{F}(\varphi)^{-1}-\mathcal{V}(\varphi),
\end{equation}
which is exactly the form of the Dirac-Born-Infeld (DBI) Lagrangian.

\section{Datasets and methodology}
\label{sec:2}
To extract the free parameters of the $\overline{\rm UTFDS}$ models, we use the recent observational datasets described below.

\textbf{Cosmic Microwave Background (CMB)}: we utilize the baseline of Planck 2018~\cite{aghanim2020planck1}, more specifically, we use the CMB temperature
and polarization angular power spectra plikTTTEEE+lowE. This base line analysis is advantageous as it avoids model-dependent non-linear effects that may introduce uncertainties.

\textbf{Redshift Space Distortions (RSD)}: RSD arises from the peculiar velocities of objects along the line of sight, leading to a mapping from real space to redshift space. This effect introduces anisotropies in the clustering patterns of objects and is influenced by the growth of cosmic structures, making RSD a sensitive probe for the combination $f\sigma_8$. In the $\overline{\rm UTFDS}$ models, $f$ is a scale-dependent quantity, which can be expressed as:
\begin{equation}\label{}
  f(k,a)=\frac{d\ln\delta(k,a)}{d\ln a}, \hspace{1cm} \delta(k,a)=\sqrt{\frac{P(k,a)}{P(k,a_0)}},
\end{equation}
where $P(k,a)$ is the matter power spectrum, which is defined by the following formula:
\begin{equation}\label{}
  \langle\widetilde{\delta}_m(\textbf{k},t)\widetilde{\delta}_m^\ast(\textbf{k}^{\prime},t)\rangle=(2\pi)^3P(\textbf{k},a(t))\delta^3(\textbf{k}-\textbf{k}^{\prime})
\end{equation}
where $\widetilde{\delta}_m(\textbf{k},t)$ is the Fourier transform of $\delta_m=\frac{\delta\rho_{\rm dm}+\delta\rho_b+\delta\rho_{\rm \overline{\nu}}}{\bar{\rho}_{\rm dm}+\bar{\rho}_b+\bar{\rho}_{\rm \overline{\nu}}}$, here the subscript $\overline{\nu}$ denotes the massive neutrino. Although dark matter behaves like a cosmological constant when the phase space trajectory close to the fixed $a$, by forcing the saddle points to lie in the future (for example, assuming $\kappa X_0\gg1$), dark matter could behave like non-relativistic matter in the present and past. $\sigma_8$ can be expressed as:
\begin{equation}\label{}
  \sigma_8(a)=\sqrt{\int_0^{+\infty}dk\frac{k^2P(k,a)W_R^2(k)}{2\pi^2}}
\end{equation}
where $W_R(k) = 3[\sin(kR)/kR-\cos(kR)]/(kR)^2$ is the Fourier transform of the top-hat window function. Here, $R$ represents the scale over which the root-mean-square (RMS) normalization of matter fluctuations is calculated.

In this study, we will utilize the RSD measurements of $f\sigma_8$ listed in Table I of Ref.~\cite{sagredo2018internal}. These measurements include 22 values of $f\sigma_8$ across the redshift range $0.02<z<1.944$, derived from various surveys: 2dFGRS~\cite{song2009reconstructing}, 2MASS~\cite{davis2011local}, SDSS-II LRGs~\cite{samushia2012interpreting}, First Amendment SNeIa+IRAS~\cite{turnbull2012cosmic,hudson2012growth}, WiggleZ~\cite{blake2012wigglez}, GAMA~\cite{blake2013galaxy}, BOSS DR11 LOWZ~\cite{sanchez2014clustering}, BOSS DR12 CMASS~\cite{chuang2016clustering}, SDSS DR7 MGS~\cite{howlett2015clustering}, SDSS DR7~\cite{feix2015growth}, FastSound~\cite{okumura2016subaru}, Supercal SNeIa+6dFGS~\cite{huterer2017testing}, VIPERS PDR-2~\cite{pezzotta2017vimos}, and eBOSS DR14 quasars~\cite{zhao2019clustering}. These measurements are commonly known as the "Gold 2018" sample in the literature. The publicly available Gold 2018 Montepython likelihood for the $f\sigma_8$
dataset is used in this article, it can be found at https://github.com/snesseris/RSD-growth. In this likelihood, the wavenumber $k$ in $f$ is fixed at 0.1 Mpc, which aligns with the effective wavenumber of the RSD measurements employed.

\textbf{Baryon Acoustic Oscillation (BAO)}: BAO analysis is derived from three sources: (1) The Six-degree Field Galaxy Survey (6dFGS)~\cite{Beutler2011The}; (2) The BAO-only portion of the eBOSS DR16 compilation~\cite{alam2021completed}, which includes data from SDSS DR7 MGS~\cite{ross2015clustering}, BOSS DR12~\cite{alam2017clustering}, eBOSS DR16 Luminous Red Galaxy (LRG) samples~\cite{bautista2021completed,gil2020completed}, eBOSS DR16 Quasar (QSO) samples~\cite{hou2021completed,hou2021completed}, eBOSS DR16 Emission Line Galaxies (ELG) samples~\cite{de2021completed}, and eBOSS DR16 Ly$\alpha$ forest samples~\cite{des2020completed}; (3) DESI year one observations as detailed in Ref.~\cite{adame2024desi1}, encompassing samples from the Bright Galaxy Sample (BGS), LRG, combined LRG and ELG, ELG, QSO, and Ly$\alpha$ forest~\cite{adame2024desi1,adame2024desi2,adame2024desi3}.

\textbf{Weak gravitational Lensing (WL)}: in addition to the above datasets, we include a prior on $S_8$, i.e.~$S_8=0.759^{+0.024}_{-0.021}$~\cite{asgari2021kids}, chosen according to the KiDS1000 measuement.
(For the $\overline{\rm UTFDS}$ models, applying the complete WL likelihood requires a detailed treatment of nonlinearities. In the absence of these tools, we limit our analysis to the linear power spectrum and assume that including the $S_8$ prior appropriately reflects the constraints from the KiDS1000 likelihoods on $\overline{\rm UTFDS}$).

To constrain the $\overline{\rm UTFDS}$ models, we run a Markov Chain Monte Carlo (MCMC) using the public MontePython-v3 code~\cite{audren2013conservative,brinckmann2019montepython} interfaced with a modified version of the CLASS code~\cite{lesgourgues2011cosmic,blas2011cosmic}. We perform the analysis with a Metropolis-Hasting algorithm and consider chains to be converged using the Gelman-Rubin~\cite{gelman1992inference} criterion $R-1<0.03$. We divided the above four sets of data into the following two datasets, namely CMB and RSD+BAO+WL, to constrain the model parameters separately. For CMB dataset, we assume flat priors on the following parameter space:
\begin{equation}\label{}
  \mathcal{P}=\{\omega_b, \Omega_{\rm dm}, \theta_s, A_s, n_s, \tau_{\rm reio}\},
\end{equation}
and for RSD+BAO+WL dataset, we assume flat priors on parameters $\Omega_{\rm dm}$, $\theta_s$, $A_s$, $n_s$, $\tau_{\rm reio}$ as well as a Gaussian prior on $\omega_b$ from Big Bang Nucleosynthesis (BBN): 100$\omega_b=2.233\pm0.036$~\cite{mossa2020baryon}.
here $\Omega_{\rm dm}=\frac{\kappa^2\rho_{\rm dm0}}{3H_0^2}$. As mentioned in the introduction, $\overline{\rm UTFDS}$ is a class of new models obtained by fixing one specific parameter in the UTFDS model, this particular parameter, is the seventh free parameter in addition to the above six basic parameters, which can be chosen as the parameter $\kappa X_0$. One of the reasons we choose to constrain $\overline{\rm UTFDS}$ instead of UTFDS is that the parameter $\kappa X_0$ can not be precisely constrained. Therefore, in this work we fix it to three discrete values, namely: 250, 500, 750, and perform a likelihood analysis on these three $\overline{\rm UTFDS}$ models. Furthermore, we adopt the Planck collaboration convention and model free-streaming neutrinos as two massless species and one massive with $M_v=0.06$~eV.

Finally, we utilize the MCEvidence code~\cite{heavens2017no,heavens2017marginal} to compute the Bayesian evidence for three $\overline{\rm UTFDS}$ models and the $\Lambda$CDM model. This is achieved by leveraging the MCMC chains used to extract the cosmological parameters. The Bayesian evidence, $B$, is calculated as the integral of the prior $\pi$ multiplied by the likelihood $L$ over the entire parameter space of the model:
\begin{equation}\label{}
  B=\int L\pi d\theta,
\end{equation}
models with larger Bayesian evidence are considered to be more supported by the data. In practice, model selection is often performed by using the logarithm of the Bayes factor $B_{ij}$, which is the logarithm of the evidence ratio between models $i$ and $j$. After computing $\ln B_{ij}$ ( where $i$ is one of the $\overline{\rm UTFDS}$ models and $j$ is the $\Lambda$CDM model) using the MCEvidence code, we apply the revised Jeffreys' scale~\cite{kass1995bayes}---an empirical scale for evaluating the strength of evidence when comparing two models, as shown in Table~\ref{tab:scale}---to quantify the evidence for three $\overline{\rm UTFDS}$ models relative to the $\Lambda$CDM model.

\begin{table}
\begin{center}
\begin{tabular}{c c }
               \hline
	            \hline
 $ \ln B_{ij}$ &  Strength of evidence for model $M_{i}$
\\ \hline
$ 0\leq\ln B_{ij}<1$    &Weak
                     \\
$1\leq \ln B_{ij}<3$&  Positive
                       \\
$3\leq \ln B_{ij}<5$&  Strong
                       \\
$ \ln B_{ij}\geq5$&  Very strong
                     \\
\hline
\hline
\end{tabular}
\end{center}
\caption{Revised Jeffreys' scale which quantifies the strength of evidence of model $M_{i}$ with respect to model $M_{j}$.}
\label{tab:scale}
\end{table}

\section{Results and discussion}
\label{sec:3}
\begin{table*}[ht]
    \centering
    \scalebox{1}[1]{
    \begin{tabular}{|c|c|c|c|c|}
        \hline
        Dataset & \multicolumn{4}{c|}{~CMB~}  \\ \hline
        Model & ~$\kappa X_0$=250~ & ~$\kappa X_0$=500~ &~$\kappa X_0$=750~ & ~$\Lambda$CDM~ \\ \hline \hline
        $100~\omega_{b}$ & $2.238\pm 0.015^{+0.030}_{-0.029} $ & $ 2.237\pm 0.015^{+0.029}_{-0.029}$ & $ 2.237\pm 0.016^{+0.032}_{-0.030}$ & $ 2.233\pm 0.015^{+0.030}_{-0.030}$ \\
        $\Omega_{\rm dm}(\Omega_{\rm cdm})$ & ${0.2676^{+0.0073}_{-0.0081}}^{+0.016}_{-0.015}$ & $0.2668\pm 0.0080^{+0.016}_{-0.016} $ & ${0.2674^{+0.0073}_{-0.0088}}^{+0.018}_{-0.017}  $ & $0.2679\pm 0.0083\pm0.016 $ \\
        $100~\theta_{s}$ & ${1.04194^{+0.00031}_{-0.00027}}^{+0.00055}_{-0.00061}$ & $1.04196\pm 0.00030^{+0.00059}_{-0.00063} $ & $1.04197\pm 0.00034^{+0.00063}_{-0.00061}$ & $1.04184\pm 0.00030^{+0.00059}_{-0.00058}  $ \\
        $\ln(10^{10}A_{s})$ & $3.050\pm 0.016^{+0.034}_{-0.033}  $ & $3.049\pm 0.018^{+0.034}_{-0.035}$ & $3.048\pm 0.017^{+0.035}_{-0.032}  $ & ${3.047^{+0.015}_{-0.017}}^{+0.033}_{-0.031}$\\
        $n_{s}$  & $0.9639\pm 0.0047^{+0.0092}_{-0.0089}$ & $0.9646\pm 0.0045^{+0.0087}_{-0.0087}$ & $0.9647\pm 0.0049^{+0.0099}_{-0.0093}$ & $0.9626\pm 0.0047^{+0.0090}_{-0.0092}$ \\
        $\tau_\mathrm{reio}$ & $0.0555\pm 0.0079^{+0.016}_{-0.015} $ & $0.0552\pm 0.0088^{+0.017}_{-0.017}$ & $0.0547\pm 0.0082^{+0.016}_{-0.016} $ & ${0.0547^{+0.0071}_{-0.0080}}^{+0.016}_{-0.015}$ \\
        $ H_0$  & $67.09\pm 0.59^{+1.2}_{-1.2}            $ & $67.12\pm 0.62^{+1.2}_{-1.2}        $ & $67.09\pm 0.74^{+1.3}_{-1.3}      $ & $67.12\pm 0.64^{+1.2}_{-1.2}      $\\
        $\Omega_{m}$& ${0.3187^{+0.0079}_{-0.0088}}^{+0.017}_{-0.016} $ & $0.3179\pm 0.0087^{+0.017}_{-0.017}          $ & ${0.3185^{+0.0081}_{-0.0091}}^{+0.019}_{-0.018}$ & $0.3189\pm 0.0091^{+0.017}_{-0.017}          $\\
        $ \sigma_8 $ & $0.7243\pm 0.0066^{+0.013}_{-0.013}          $ & $0.7875\pm 0.0079^{+0.015}_{-0.015} $ & ${0.8013^{+0.0073}_{-0.0081}}^{+0.015}_{-0.015} $ & $0.8129\pm 0.0078^{+0.016}_{-0.015}          $\\
        $ S_8$ & $0.747\pm 0.014^{+0.027}_{-0.028}           $ & $0.811\pm 0.016^{+0.032}_{-0.032}            $ & $0.826\pm 0.020^{+0.035}_{-0.034}            $ &      $0.838\pm0.017\pm0.033 $                       \\ \hline
        $\chi_{\rm min}^2$ &2750.12 &2748.34 & 2748.66 & 2747.06 \\
        $\Delta\chi_{\rm min}^2$ &$ 3.06$& $1.28 $&$1.60 $&$ 0 $\\
        \hline
    \end{tabular}}
    \caption{The mean values and 1, 2$\sigma$ errors of the parameters as well as minimum chi-square value of three $\overline{\rm UTFDS}$ models with the $\Lambda$CDM model for the CMB dataset.}
    \label{tab:1}
\end{table*}

\begin{table*}[ht]
    \centering
    \scalebox{1}[1]{
    \begin{tabular}{|c|c|c|c|c|}
        \hline
        Dataset & \multicolumn{4}{c|}{~RSD+BAO+WL~}  \\ \hline
        Model & ~$\kappa X_0$=250~ & ~$\kappa X_0$=500~ &~$\kappa X_0$=750~ & ~$\Lambda$CDM~ \\ \hline \hline
        $100~\omega_{b }$ & $2.251\pm 0.037^{+0.069}_{-0.071}            $ & $2.232\pm 0.035^{+0.069}_{-0.071}            $ & $2.234\pm 0.035^{+0.069}_{-0.069}            $& $            2.235\pm 0.037^{+0.073}_{-0.071}$\\
        $\Omega_{\rm dm }(\Omega_{\rm cdm})   $ & $0.266\pm 0.012^{+0.024}_{-0.023}            $ & $0.248\pm 0.011^{+0.021}_{-0.020}             $ & $0.246\pm 0.010^{+0.021}_{-0.020}            $ & $    0.248\pm 0.010^{+0.020}_{-0.020}       $                         \\
        $H_0            $ & $69.04\pm 0.63^{+1.3}_{-1.2}             $ & $68.22\pm 0.60^{+1.2}_{-1.1}             $ & ${68.24^{+0.56}_{-0.63}}^{+1.2}_{-1.1}        $ & $  68.25\pm 0.61^{+1.2}_{-1.1}           $ \\
        $\Omega_{m }  $ & $0.315\pm 0.012^{+0.023}_{-0.023}            $ & $0.297\pm 0.010^{+0.021}_{-0.020}            $ & $0.296\pm 0.010^{+0.020}_{-0.020}            $  & $ 0.2970\pm 0.0098^{+0.019}_{-0.019}            $\\
        $   \sigma_8       $ & $0.772\pm 0.023^{+0.049}_{-0.046}            $ & $0.778\pm 0.021^{+0.043}_{-0.040}             $ & $0.774\pm 0.021^{+0.041}_{-0.039}            $ & $    0.772\pm 0.029^{+0.056}_{-0.057}      $ \\
        $ S_8            $   & $0.790\pm 0.019^{+0.036}_{-0.036}             $ & $0.774\pm 0.019^{+0.037}_{-0.036}             $ & $0.768\pm 0.018^{+0.035}_{-0.034}             $ & $  0.768\pm 0.029^{+0.058}_{-0.057}    $                        \\
        \hline
        $\chi_{\rm min}^2$ & 89.48  & 40.44  & 38.86  & 39.22  \\
        $\Delta\chi_{\rm min}^2$ &$ 50.26 $& $ 1.22  $&$-0.36 $&$ 0 $\\
        \hline
    \end{tabular}}
    \caption{The mean values and 1, 2$\sigma$ errors of the parameters as well as minimum chi-square value of three $\overline{\rm UTFDS}$ models with the $\Lambda$CDM model for the RSD+BAO+WL dataset.}
    \label{tab:2}
\end{table*}

\begin{table*}[ht]
    \centering
    \scalebox{1}[1]{
    \begin{tabular}{|c|c|c|c|c|}
        \hline
        Model & ~$\kappa X_0$=250~ & ~$\kappa X_0$=500~ &~$\kappa X_0$=750~ & ~$\Lambda$CDM~ \\ \hline \hline
        $ \sigma_8 $ tension  & 2.0$\sigma$   &  0.4$\sigma$    &  1.2$\sigma$    & 1.4$\sigma$    \\
        $ S_8 $ tension  & 1.8$\sigma$   &  1.5$\sigma$    &  2.2$\sigma$    & 2.1$\sigma$    \\
        \hline
    \end{tabular}}
    \caption{The $\sigma_8$ and $S_8$ tensions between datasets CMB and RSD+BAO+WL for three $\overline{\rm UTFDS}$ models and the $\Lambda$CDM model.}
    \label{tab:3}
\end{table*}

\begin{table*}[ht]
    \centering
    \scalebox{1}[1]{
        \begin{tabular}{|c| c| c| c|}
        \hline
            Model                  & Datasets               & $\ln B_{ij}$& Strength of evidence for $\Lambda$CDM \\
            \hline\hline
            $\kappa X_0$=250                   & CMB                    & $-1.46$ &  Definite/Positive         \\
            $\kappa X_0$=500                   & CMB                & $-1.11$ &  Definite/Positive       \\
            $\kappa X_0$=750                   & CMB       & $-1.56$   &  Definite/Positive     \\
            \hline
            $\kappa X_0$=250 & RSD+BAO+WL                    & $-24.65$ &  Very strong        \\
            $\kappa X_0$=500 & RSD+BAO+WL                & $-1.57$  & Definite/Positive      \\
            $\kappa X_0$=750 & RSD+BAO+WL       & $-0.827$  &  Weak     \\ \hline
        \end{tabular}}
    \caption{Summary of the $\ln B_{ij}$ values quantifying the evidence of fit for three $\overline{\rm UTFDS}$ models relative to the $\Lambda$CDM model for datasets CMB and RSD+BAO+WL. One note that a negative value of $\ln B_{ij}$ indicates that the $\overline{\rm UTFDS}$ models is less supported compared to the base model $\Lambda$CDM.}
    \label{tab:lnB}
\end{table*}

\begin{figure*}
	\centering
	\includegraphics[scale=0.5]{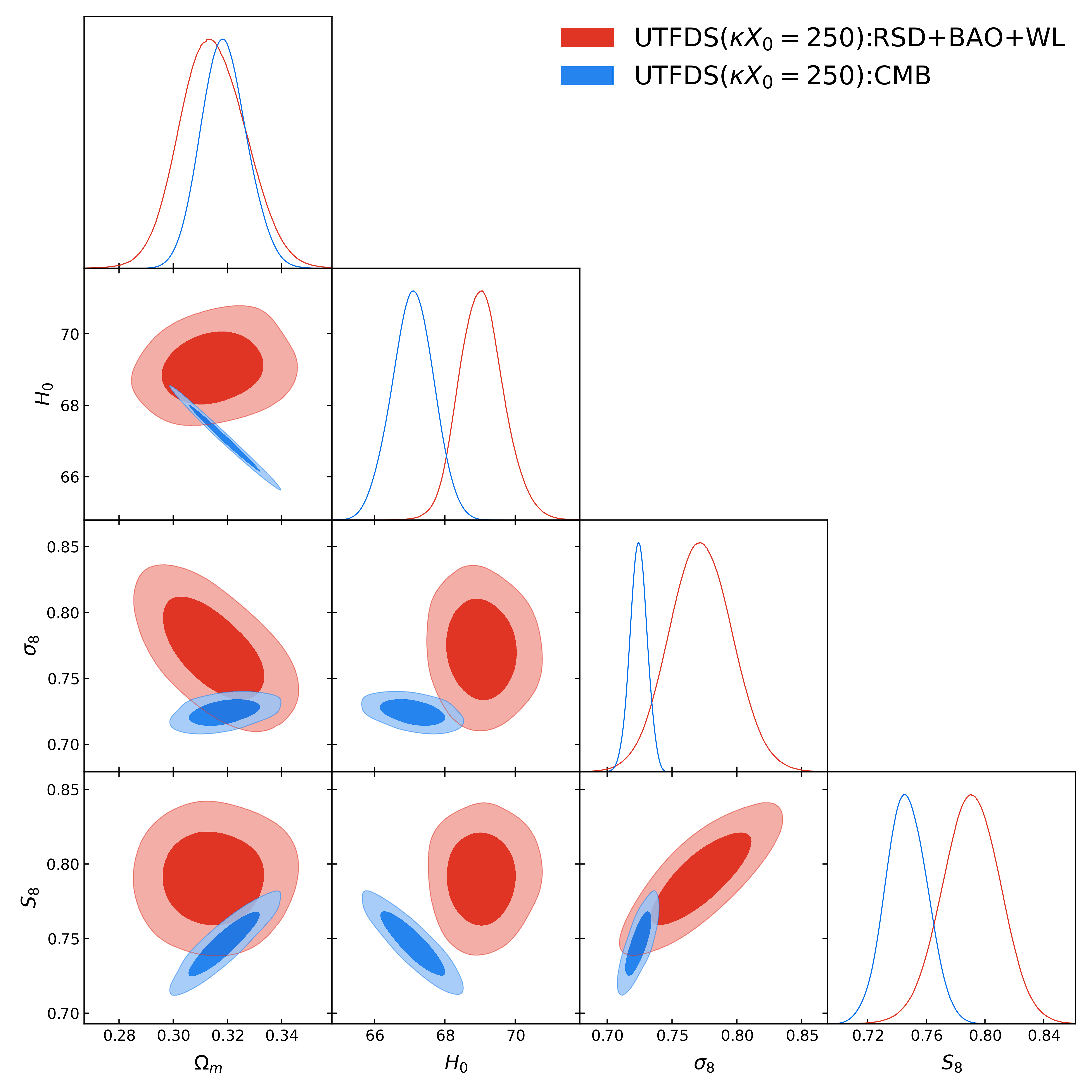}
	\caption{One dimensional posterior distributions and two dimensional joint contours at 68\% and 95\% CL for the most relevant parameters of the $\overline{\rm UTFDS}$ model with $\kappa X_0$=250 using CMB and RSD+BAO+WL datasets.}\label{fig:1}
\end{figure*}

\begin{figure*}
	\centering
	\includegraphics[scale=0.5]{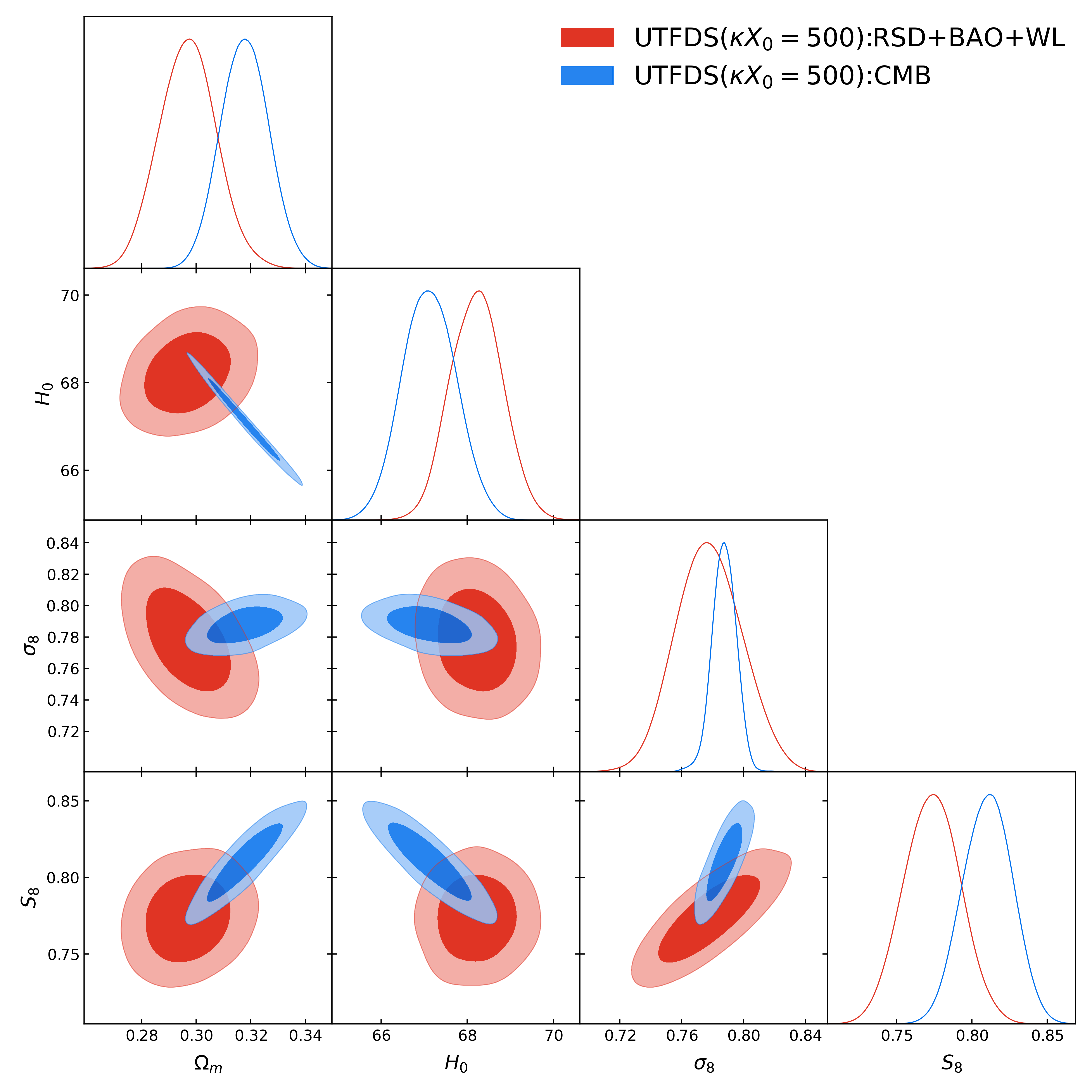}
	\caption{One dimensional posterior distributions and two dimensional joint contours at 68\% and 95\% CL for the most relevant parameters of the $\overline{\rm UTFDS}$ model with $\kappa X_0$=500 using CMB and RSD+BAO+WL datasets.}\label{fig:2}
\end{figure*}

\begin{figure*}
	\centering
	\includegraphics[scale=0.5]{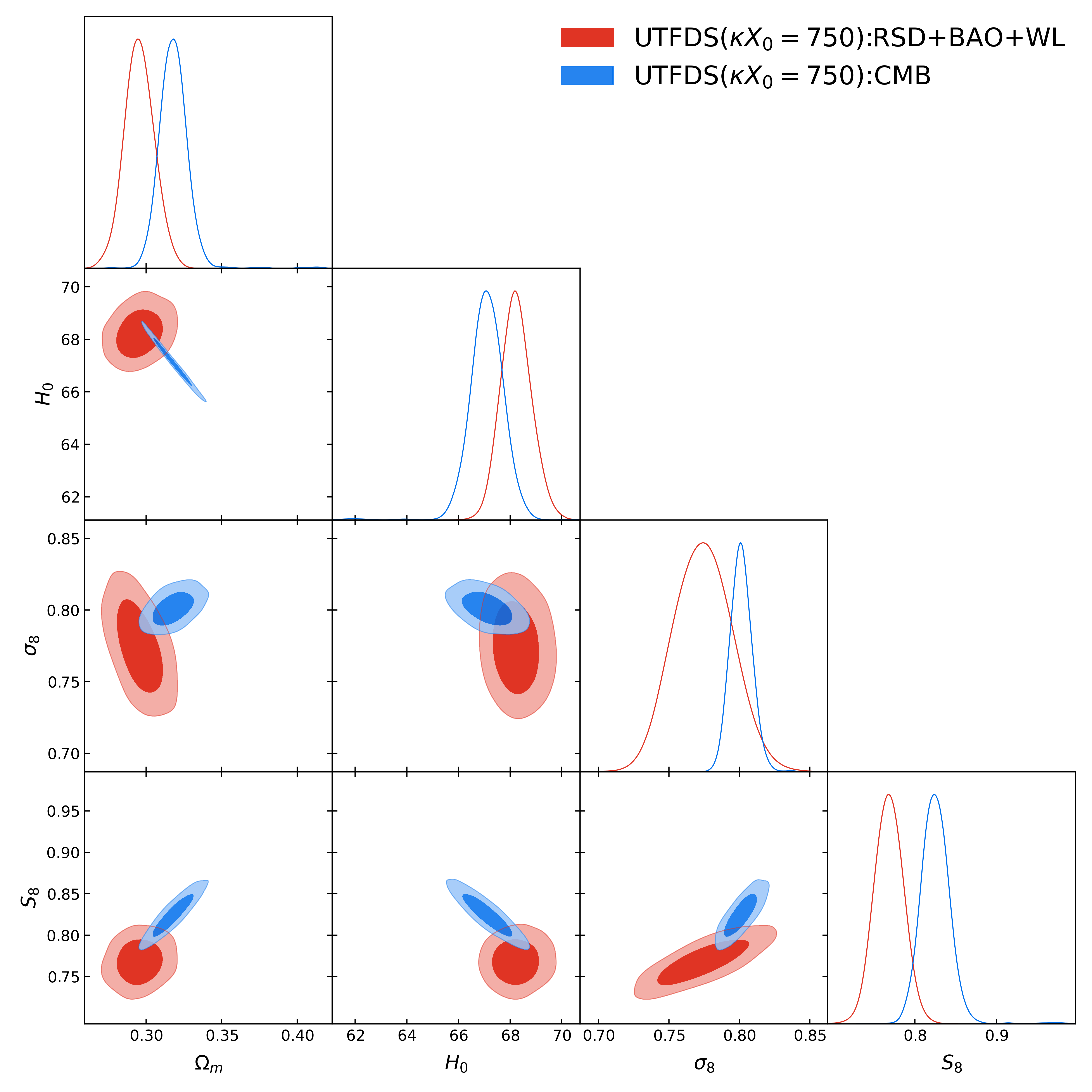}
	\caption{One dimensional posterior distributions and two dimensional joint contours at 68\% and 95\% CL for the most relevant parameters of the $\overline{\rm UTFDS}$ model with $\kappa X_0$=750 using CMB and RSD+BAO+WL datasets.}\label{fig:3}
\end{figure*}

\begin{figure*}
	\centering
	\includegraphics[scale=0.5]{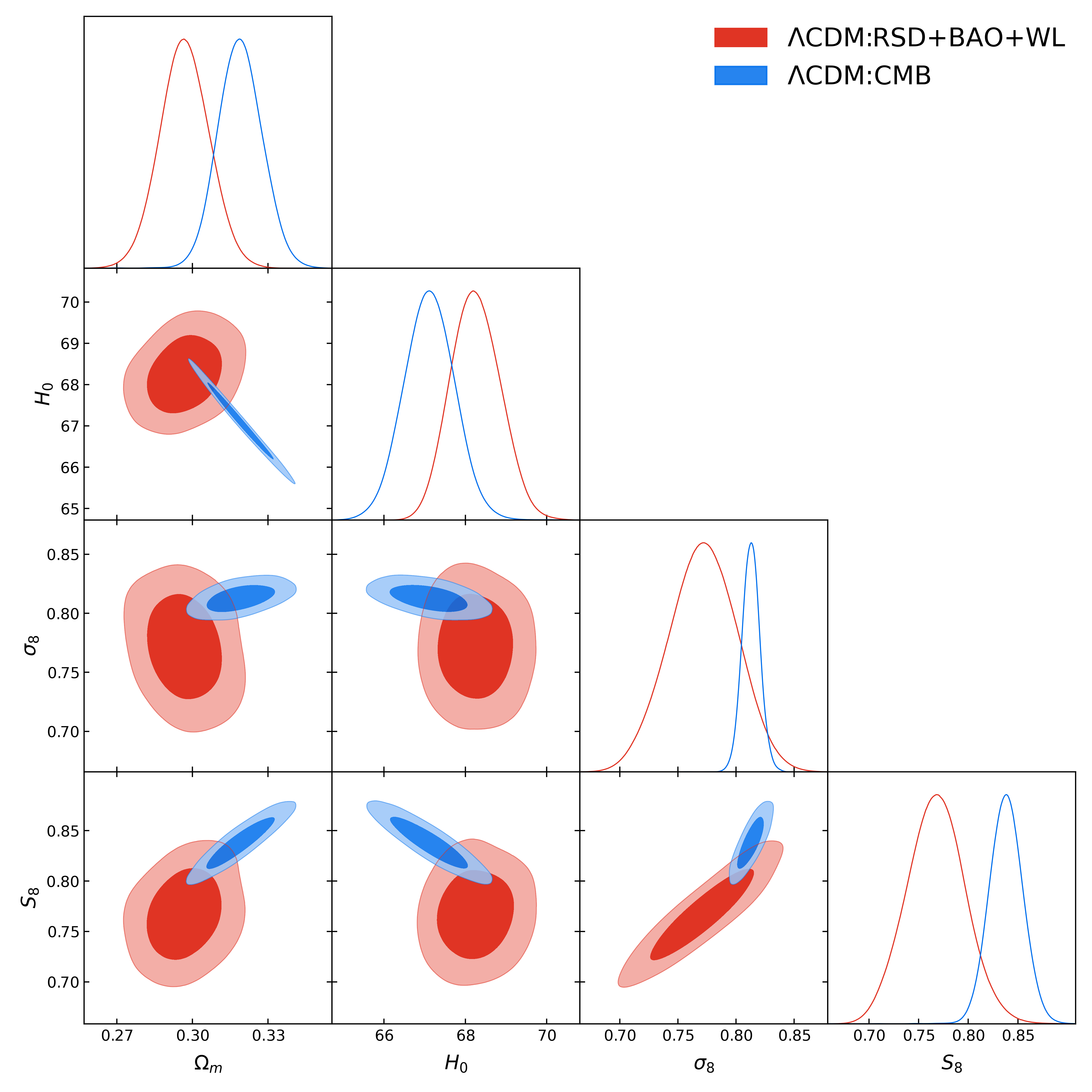}
	\caption{One dimensional posterior distributions and two dimensional joint contours at 68\% and 95\% CL for the most relevant parameters of the $\Lambda$CDM model using CMB and RSD+BAO+WL datasets.}\label{fig:4}
\end{figure*}
In Tab.~\ref{tab:1}-\ref{tab:2}, we present the constraints on three $\overline{\rm UTFDS}$ models and the $\Lambda$CDM model for CMB and RSD+BAO+WL datasets. And Fig.~\ref{fig:1}-\ref{fig:4} show the one dimensional posterior distributions and two dimensional joint contours at 68\% and 95\% confidence levels for the most relevant parameters of three $\overline{\rm UTFDS}$ models and the $\Lambda$CDM model.

From Tab.~\ref{tab:1}, one can see that except for parameters $\sigma_8$ and $S_8$, the fitting results for the other parameters in these four models show little difference. Also, it can be inferred that there is a positive correlation between the parameter $\kappa X_0$ and the parameters $\sigma_8$ and $S_8$. Recalling that the dark matter sound speed squared is given by $c_{(\rm dm)s}^2=\frac{1}{1+\frac{2}{3}\tan^2{\frac{\pi x}{2}}}=\frac{1}{1+\kappa^2X^2}$, and solving Eq.~34 under the assumption $\kappa X\gg y$, we find that $\kappa X$ behaves as $\kappa X\sim\kappa X_0(a_0/a)^{6}$, this implies that the larger $\kappa X_0$ is, the smaller $c_{(\rm dm)s}^2$ becomes, which makes intuitive sense, as a larger dark matter sound speed corresponds to a smoother Universe, that is, a Universe with smaller values of $\sigma_8$ and $S_8$. Furthermore, based on the minimum chi-square values of these four models, the $\Lambda$CDM model has the smallest minimum chi-square value, followed by the $\overline{\rm UTFDS}$ model with $\kappa X_0$=500, 750, and 250, with their minimum chi-square values being 1.28, 1.60, and 3.06 higher than that of the $\Lambda$CDM model, respectively.

From Tab.~\ref{tab:2}, one can see that there is no longer a positive correlation between the parameter $\kappa X_0$ and parameters $\sigma_8$ and $S_8$. This is because, in the previous situation, parameter $\kappa X_0$ was only correlated with parameters $\sigma_8$ and $S_8$ and not with the other parameters, as evidenced by the fact that the fitting results for the other parameters remained nearly unchanged when parameter $\kappa X_0$ took different values. However, in the current situation, parameter $\kappa X_0$ is not only correlated with parameters $\sigma_8$ and $S_8$ but also with the other parameters. This leads to different degrees of variation in the fitting results for the other parameters when parameter $\kappa X_0$ takes different values. Since these other parameters are themselves correlated with parameters $\sigma_8$ and $S_8$, this ultimately results in parameters $\sigma_8$ and $S_8$ no longer increasing as parameter $\kappa X_0$ increases. In addition, given the minimum chi-square values of these four models, one finds that the $\overline{\rm UTFDS}$ model with $\kappa X_0$=750 has the smallest minimum chi-square value, followed by the the $\Lambda$CDM model, the $\overline{\rm UTFDS}$ model with $\kappa X_0$=500, 250, with their minimum chi-square values being 0.36, 1.58, and 50.62 higher than that of the $\overline{\rm UTFDS}$ model with $\kappa X_0$=750, respectively. It is clear that the $\overline{\rm UTFDS}$ model with $\kappa X_0$=250 is strongly excluded by RSD+BAO+WL dataset.

Now let's examine how these three $\overline{\rm UTFDS}$ models perform in alleviating the $\sigma_8$ and $S_8$. From Tab.~\ref{tab:1}, one finds that for CMB dataset, we have ($\sigma_8=0.7243\pm 0.0066$, $S_8=0.747\pm0.014$), ($\sigma_8=0.7875\pm 0.0079$, $S_8=0.811\pm 0.016$), ($\sigma_8=0.8013^{+0.0073}_{-0.0081}$, $S_8=0.826\pm 0.020$), and ($\sigma_8=0.8129\pm0.0078$, $S_8=0.838\pm0.017$) for the $\overline{\rm UTFDS}$ model with $\kappa X_0$=250, 500, 750, and the $\Lambda$CDM model, respectively. And from Tab.~\ref{tab:2}, one finds that for the RSD+BAO+WL dataset, we have ($\sigma_8=0.772\pm 0.023$, $S_8=0.790\pm 0.019$), ($\sigma_8=0.778\pm 0.021$, $S_8=0.774\pm 0.019$), ($\sigma_8=0.774\pm 0.021$, $S_8=0.768\pm 0.018$), and ($\sigma_8=0.772\pm 0.029$, $S_8=0.768\pm 0.029$) for the $\overline{\rm UTFDS}$ model with $\kappa X_0$=250, 500, 750, and the $\Lambda$CDM model, respectively. Therefore, as is shown in Tab.~\ref{tab:3}, the $\sigma_8$($S_8$) tension between datasets CMB and RSD+BAO+WL for the $\overline{\rm UTFDS}$ model with $\kappa X_0$=250, 500, 750, and the $\Lambda$CDM model are  2.0$\sigma$(1.8$\sigma$), 0.4$\sigma$(1.5$\sigma$), 1.2$\sigma$(2.2$\sigma$) and 1.4$\sigma$(2.1$\sigma$). One notes that the $\overline{\rm UTFDS}$ model with $\kappa X_0$=500 is the only one among the four models where both $\sigma_8$ and $S_8$ tensions are below 2.0$\sigma$, it is also worth noting that the tensions are relieved without exacerbating the Hubble tension. The $\sigma_8$ and $S_8$ tensions can be intuitively seen from Fig.~\ref{fig:1}-\ref{fig:4}.

Finally, we present the $\ln B_{ij}$ values quantifying the evidence of fit for three $\overline{\rm UTFDS}$ models relative to the $\Lambda$CDM model for datasets CMB and RSD+BAO+WL, which is shown in Tab.~\ref{tab:lnB}. Recalling the Revised Jeffreys' scale shown in Tab.~\ref{tab:scale}, we find that the CMB dataset favors $\Lambda$CDM over all of three $\overline{\rm UTFDS}$ models with definite/positive evidence. And the RSD+BAO+WL dataset favors $\Lambda$CDM over the $\overline{\rm UTFDS}$ model with $\kappa X_0$=250, 500, and 750 with very strong, definite/positive, and weak evidences. Therefore, although the $\overline{\rm UTFDS}$ model with $\kappa X_0$=500 performs better than $\Lambda$CDM when it comes to alleviating both $\sigma_8$ and $S_8$ tensions, both datasets CMB and RSD+BAO+WL favor $\Lambda$CDM over the $\overline{\rm UTFDS}$ model with $\kappa X_0$=500 with definite/positive evidence. However, since this evidence is not strong, the $\overline{\rm UTFDS}$ model with $\kappa X_0$=500 is still worth further investigation.

\section{concluding remarks}
With advances in observational technology, astronomers have been able to constrain cosmological parameters with increasing precision. This has led to inconsistencies in some parameters derived from different datasets within the $\Lambda$CDM model, with the most notable discrepancies being the $H_0$ tension and the $\sigma_8$ (or $S_8$) tension. In order to relieve the latter, in this paper, we have put forward a unified dark sector model, the UTFDS model, which unifies dark energy and dark matter using a three-form field. Here, the potential of the three-form field serves as dark matter, and the kinetic term represents dark energy. The interaction between dark matter and dark energy arises from the energy exchange between these two terms.

We have not only presented the dynamical equations of UTFDS at both the background and linear levels, but also derived the autonomous system of evolution equations for UTFDS and conducted a stability analysis of its fixed points. The analysis results suggest that, in the absence of radiation and baryons and at least at background level, UTFDS acts like dust matter at high redshifts, while behaves like a cosmological constant at low redshifts, aligning with our expectations for a unified dark sector. We have also found that the dual Lagrangian of the UTFDS Lagrangian is equivalent to a DBI Lagrangian.

Since the parameter $\kappa X_0$ can not be precisely constrained, in this work we have fixed it to three discrete values, namely: 250, 500, 750. And we have referred to the resulting models as the $\overline{\rm UTFDS}$ model with $\kappa X_0$=250, 500, and 750. Then we have constrained these three models and the $\Lambda$CDM model separately using the CMB and RSD+BAO+WL datasets, and found that for the CMB dataset, there is a positive correlation between the parameter $\kappa X_0$ and the parameters $\sigma_8$ and $S_8$, which is result from the negative correlation between the sound speed of dark matter and the parameters $\sigma_8$ and $S_8$. For the RSD+BAO+WL dataset, this correlation was not found, we attributed the reason for this to that there are different degrees of variation in the fitting results for the basic parameters when parameter $\kappa X_0$ takes different values and these parameters are themselves correlated with parameters $\sigma_8$ and $S_8$. By calculating the $\sigma_8$ and $S_8$ tensions between datasets CMB and RSD+BAO+WL for the $\overline{\rm UTFDS}$ model with $\kappa X_0$=250, 500, 750, and the $\Lambda$CDM model, we have found that the $\overline{\rm UTFDS}$ model with $\kappa X_0$=500 is the only one among the four models where both $\sigma_8$ and $S_8$ tensions are below 2.0$\sigma$, furthermore, the tensions are relieved without exacerbating the Hubble tension. However, although the $\overline{\rm UTFDS}$ model with $\kappa X_0$=500 performs better than $\Lambda$CDM in addressing both $\sigma_8$ and $S_8$ tensions, both CMB and RSD+BAO+WL datasets provide definite/positive evidence in favor of $\Lambda$CDM over the $\overline{\rm UTFDS}$ model with $\kappa X_0$=500. Nevertheless, due to the evidence being not strong, the $\overline{\rm UTFDS}$ model with $\kappa X_0$=500 still warrants further investigation.
\section*{Acknowledgments}
This work is supported by Guangdong Basic and Applied Basic Research Foundation (Grant No.2024A1515012573), the National key R\&D Program of China (Grant No.2020YFC2201600),  National Natural Science Foundation of China (Grant No.12073088), and National SKA Program of China (Grant No. 2020SKA0110402).

\bibliographystyle{spphys}
\bibliography{unified}

\end{document}